\documentclass{scrartcl}       

\usepackage{booktabs} 
\usepackage{url}
\usepackage{amssymb}
\usepackage{graphicx}
\usepackage{subfigure}
\usepackage{color}
\usepackage{csquotes}
\usepackage{setspace}
\usepackage{graphicx}
\usepackage{subfigure}
\usepackage{color}
\usepackage{algorithm}
\usepackage{algorithmic}
\usepackage[hidelinks]{hyperref} 
\usepackage[affil-it]{authblk}

\usepackage{proof}

\usepackage{amsthm}
\usepackage{stmaryrd}

\theoremstyle{definition} 
\newtheorem{Definition}{Definition}

\usepackage{mathptmx}      

\begin{document}

\title{Transformation Rules for the decentralization of a blockchain-extended global process model}
\subtitle{Technical Report, Version 0.1}

\author{Julius K\"{o}pke\thanks{Electronic address: \texttt{julius.koepke@aau.at}; Corresponding author } }
\author{Sebastian Trattnig}
\affil{Institute for Informatics Systems, Universität Klagenfurt}

\date{\today}

\maketitle

\begin{abstract}
Blockchains and distributed ledger technology offer promising capabilities for supporting collaborative business processes across organizations. 
Typically, approaches in this field fall into two categories: either executing the entire process model on the blockchain or using the blockchain primarily to enforce or monitor the exchange of messages between participants. Our work proposes a novel approach that sits between these two methods.

We introduce a centralized process model extended with blockchain annotations, detailing the tasks of each participating organization and the extent to which block\-chain technology is needed to secure task execution. This model also includes all critical data objects and specifies how their handling should be protected by the blockchain.

This technical report outlines a systematic three-step method for automatically decentralizing this comprehensive model into individual local process models for each organization, coupled with a separate process model for the blockchain. This decentralized structure effectively replicates the original global process model.

Our transformation approach is rule-based, focusing on creating a platform-inde\-pendent model first, then a platform-specific model. Subsequently, we project the platform-specific model to obtain one model for the blockchain and one model for each participating organization.

\end{abstract}

\section{Introduction}
\begin{figure}[H]
\center
\includegraphics[width=\textwidth]{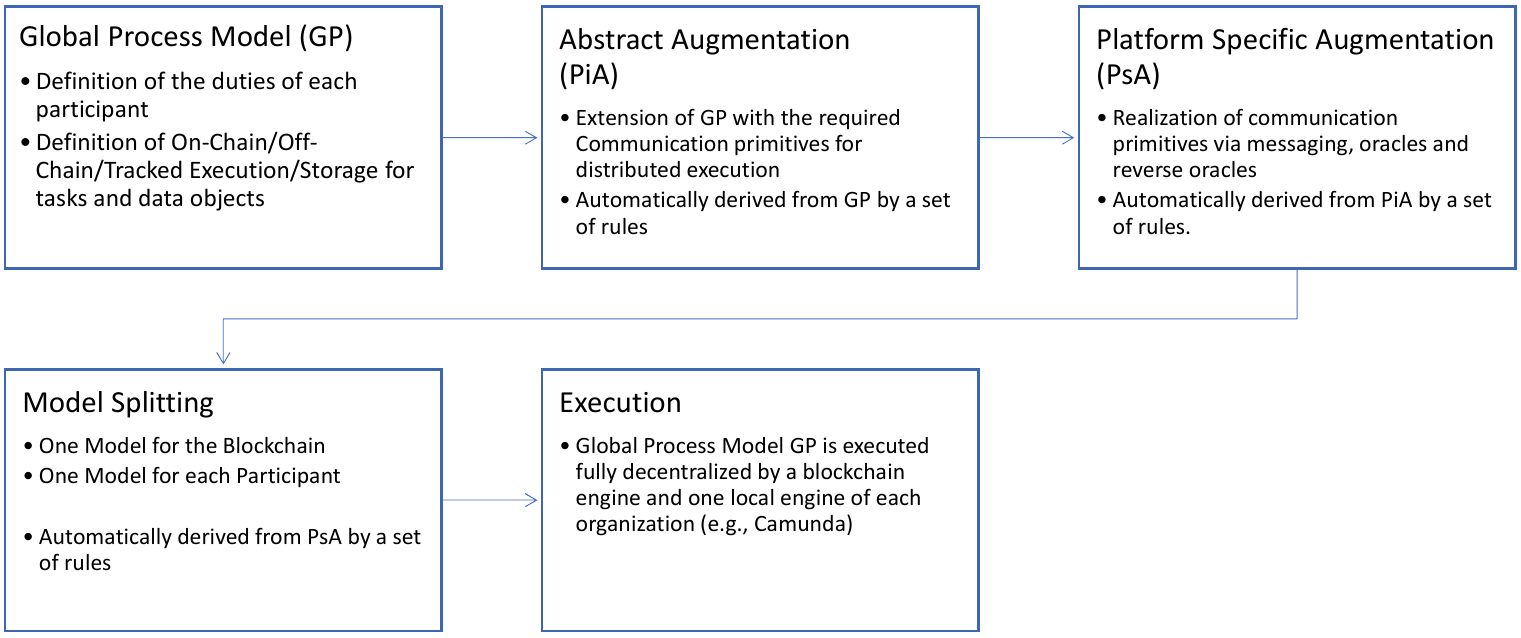}
\caption{General Overview}
\label{fig:overview}
\end{figure}

This technical report describes a method for transforming a blockchain-extended global process model into multiple Business Process Models. Each model is designated to be executed on an organization's process engine (such as Camunda BPMN), and one model is designed to be executed on a Blockchain (such as Ethereum or Hyperledger). 
The process begins with a global process model (GP), which contains standard process meta-model features. These features include tasks, gateways, loops, data objects, control-flow edges, and data-flow edges. The model is inter-organizational, as each task or gateway is assigned to the relevant organization. The model is also blockchain-extended because blockchain-specific properties are added to process elements. Tasks can be defined as on-chain, off-chain, or tracked, and data objects can be on-chain, off-chain, or digest-on-chain. Figure \ref{fig:overview} provides an overview of the process.

After obtaining a model as input, we apply a set of rules to derive a Platform Independent Augmentation (PiA), which includes all the necessary communication between different organizations. However, this model lacks details on how the communication should be implemented.

The PiA serves as input for the next transformation step where platform-specific communications replace abstract ones to obtain a platform-specific augmentation. For instance, if the communication is completely off-chain, classical message exchanges are introduced. Oracles \cite{oracleFoundation} are included for communication from an off-chain model to the on-chain Model, whereas reverse oracles are used for communication from the blockchain model to off-chain models. Additionally, implementation-specific exchanges of data objects between the different models are added based on the blockchain properties of the data objects.

Finally, the resulting global Platform Specific Augmentation is split into one model for each participant and one model for execution on the blockchain.

The paper is structured as follows: In Section \ref{sect:globalmodel}, we present the Meta-Model for Global Processes, which is based on an extended subset of BPMN2 process diagrams. Section \ref{sect:RulesPiA} introduces the rules for transforming this model to obtain a platform-inde\-pendent Augmentation. We then present the generation of a platform-specific augmentation that covers the control-flow aspect in Section \ref{sec:genPisa}. In Section \ref{sect:dataAccess}, we introduce the rules for implementing general data access within PsA. Finally, we describe in Section \ref{sect:projection} how one local process model for each organization and one process model for the blockchain are derived from the PsA.

\section{Global Process Meta Model} \label{sect:globalmodel}
The immutability property of blockchains imposes additional requirements on the correctness of process models. A design error in a deployed process model, which is, in turn, realized by a smart contract, can typically not be modified after deployment. As a consequence, instances may get stuck forever with no relief. There is, by purpose, no administrator who can cancel a case or apply ad-hoc changes. The possibility that funds in cryptocurrencies are locked forever due to a stuck process further emphasizes the role of correctness. Based on these assumptions, we use a reduced - still expressive - process meta-model that allows correctness by design. We support the core control-flow patterns sequence, xor, par, and structured loops.
We restrict process models to the full-blocked \cite{wfmc_xpdl} class defined by the WfMC. Therefore, every split node must have a corresponding join node.
Models following these restrictions are by design sound. \\

We base our meta-model on the one of \cite{koepkeiiwas14} and its extension with data from \cite{Köpke2019}. To allow direct implementations of the model as a subset of BPMN, we adopt the originally block-based model to a graph-based representation. Furthermore, we extend it with blockchain-specific properties proposed in SecBPMN2BC \cite{secbpmn2bc}. In particular, we add the following blockchain-specific properties:

\begin{itemize}
\item Activities can be defined as \emph{on-chain, off-chain}, or \emph{tracked}. An on-chain activity is triggered by the respective actor, but the task itself is realized as smart contract code. Off-chain tasks are executed on some process engine. In contrast, tracked tasks are executed off-chain, but their start and end events are recorded on-chain. 

\item Data objects can be marked as \emph{on-chain}, \emph{off-chain}, or \emph{digest}. An on-chain data object is stored on the blockchain. An off-chain data object is stored off-chain by the respective process engines. A data object with the storage model \emph{digest} is stored off-chain, but the hash of the data values and the last actor writing to the data object is recorded on-chain. 
\end{itemize}
The additional actor `chain` defines that the blockchain enforces the coordination of a control structure (xor/par/loop).

We follow an explicit approach for modeling decisions in processes using business rule tasks. This allows a clear definition of the actors in charge of making decisions and informing dependent actors. It also allows a straightforward implementation of the models using DMN \cite{dmnpaper} decision tables. We require that each xor-gateway has a business rule task as its direct predecessor and each loop body ends with a business rule task for computing the re-execution criteria.

\begin{Definition}[Global Process Model]\label{def:global_process_model} 
A global process model is a directed acyclic graph $P = (N,$ $E,$ $O,$ $D,$ $A)$, where nodes in $N$ represent activities, start-events, end-events, xor-split/join, pars-split/join, and (structured) loops. Each node $n \in N$ has the attribute $n.type$ to define the node type (See Def \ref{def:node_types_global} for details).
Each edge $e \in E$ has the form $e=(n1,n2,l)$ and defines the precedence relation between $n1$ and $n2$. Therefore, $n1$ is a direct predecessor of $n2$ and $n2$ is a direct successor of $n1$, $l$ is an optional label. 

$O$ is a set of data objects, $D \subset O$ is a set of decision variables used to control xor gateways and loops. $A$ is a set of actors. $A$ contains the special actor $`chain`$ (see above). We use the object-style notation. E.g., $P.A$ is the set of actors of the process $P$. Each node $n \in N$ has an actor $n.a \in P.A$ responsible for its execution. 
Each data object in $O \setminus D$ has the property $d.s \in \{`offchain`, `onchain`, `digest`\}$ to define its storage model (see above).

The process model is full-blocked. Therefore, each xor- and par-split node must have a corresponding join node.
\end{Definition}

\begin{Definition}[Node Types of global process Model]\label{def:node_types_global}
Let $P$ be a global process model. Each $n \in P.N$ can have one of the following types.
\begin{itemize}
\item \underline{Activity}, ($n.type=`Activity`$). An activity has a label $n.label$, a read set defining all input data objects $n.r\subseteq P.O$ and a write set $n.w\subseteq P.O$. These are used to define the implicit data flow between activities.   

To define whether an activity is executed fully off-chain, only tracked on the blockchain, or fully on-chain, activities have the additional property $n.exec \in \{`offChain`, `tracked`, `onChain`\}$ (see above).

Constraints: For each activity $n$ in $P.N$, there is exactly one incoming edge $(\_,n,\_) \in P.E$ and exactly one outgoing edge $(n,\_,\_) \in P.E$.  

\item \underline{PAR-Split}, ($n.type=`ParSplit`$). Par splits have the usual semantics of parallel split gateways. 

Constraints: For each par-split node $n$ in $P.N$, there are exactly two outgoing edges of the form $(n,\_,\_) \in P.N$ and exactly one incoming edge $(\_,n,\_) \in P.E$.  

\item \underline{PAR-Join}, ($n.type=`ParJoin`$). Par Joins have the usual semantics of parallel join gateways. 

Constraints: For each par-join node $n$ in $P.N$, there is exactly one outgoing edge of the form $(n,\_,\_) \in P.E$ and exactly two incoming edges of the form $(\_,n,\_) \in P.E$.  
\item \underline{Start}, ($n.type=`Start`$). Start nodes represent the start event of the process.

Constraints: For each start node $n$ in $P.N$, there is exactly one outgoing edge of the form $(n,\_,\_) \in P.E$, and there must be no incoming edge $(\_,n,\_) \in P.E$.  

\item \underline{End}, ($n.type=`End`$).  End nodes represent the end event of the process.

Constraints: For each end node $n$ in $P.N$, there is exactly one incoming edge of the form $(\_, n,\_) \in P.E$, and there must be no edge outgoing $(n,\_,\_) \in P.E$.

\item \underline{Business Rule Task}, ($n.type=`BusinessRuleTask`$). Business rule tasks derive the decision variable for xor-splits and loops.
Such tasks are special activities where the write set $n.w=\{c\}$, with $c \in P.D$.

Constraints: Each decision variable in $P.D$ is assigned to exactly one Business Rule Task in $P.N$.

\item \underline{XOR-Split}, ($n.type=`XorSplit`$). Nodes of type $xorSplit$ have the additional property $n.c \in P.D$. If the decision variable has the value $true$ at runtime, the edge $(n,\_,n.c) \in P.E$ is activated, otherwise the edge  $(n,\_,!n.c)$.

Constraints: For each xor-split node $n$ in $P.N$, there are exactly two outgoing edges of the form $(n,\_,\_) \in P.E$.
In particular, there is exactly one outgoing edge of the form $(n,\_,n.c) \in P.E$ and exactly one edge of the form $(n,\_,!n.c) \in P.E$. 
Additionally, there is exactly one incoming edge of the form  $(\_,n,\_) \in P.E$. In particular, this edge must have the form $(b,n,\_)$, where $b.type=`businessRuleTask`$ and $b.w=\{n.c\}$.

\item \underline{XOR-Join}, ($n.type=`XorJoin`$) XOR-Joins have the usual semantics of xor join gateways. 

Constraints: For each xor-join node $n$ in $P.N$, there is exactly one outgoing edge of the form $(n,\_,\_) \in P.E$ and exactly two incoming edges of the form $(\_,n,\_) \in P.E$. 

\item \underline{Sub Process}, ($n.type=`SubProcess`$). Loops are supported via nodes of type $subProcess$ with the additional property $n.body$ and $n.loopCondition$. Property $n.body$ refers to a process model $PL$, where $PL.D \subseteq P.D$, $PL.O \subseteq P.O$  and $PL.A \subseteq P.A$, and 
$n.loopCondition$ is linked to a decision variable $\in$ $P.D$. If the variable holds the value $true$ after executing the sub process, it is repeated. 

Constraints: For each sub process $n$ in $P.N$, there is exactly one incoming edge $(\_,n,\_) \in P.E$ and exactly one outgoing edge $(n,\_,\_) \in P.E$.\\
If $n.loopCondition$ is set, there must be an edge $(b,e,\_) \in n.body.E$ such that $b.type=`BusinessRuleTask`$ and $e.type=`End`$, and $n.loopCondition \in b.w$
\end{itemize}

\end{Definition}

\subsection{Graphical Notation}
\begin{figure}[H]
\center
\includegraphics[width=\textwidth]{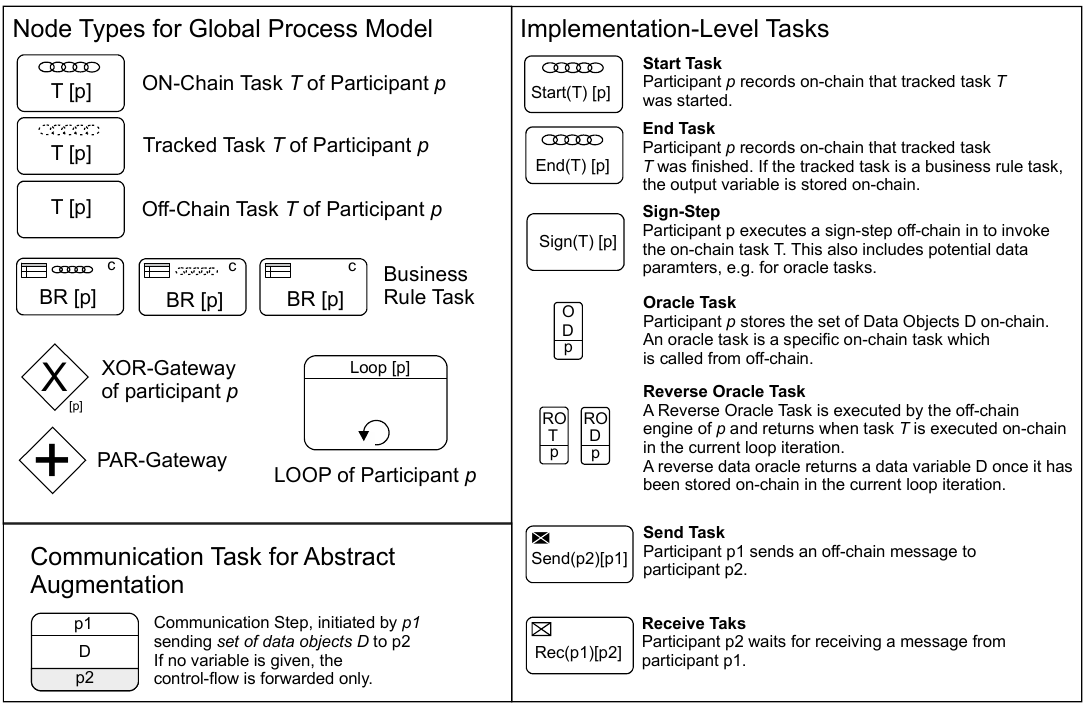}
\caption{Node Types of Global Process Model and Augmentations}
\label{GlobalprocessModel}
\end{figure}

We use a BPMN-based graphical notation for our process meta-model shown in the left upper part of Figure \ref{GlobalprocessModel}.
The actors of a node $p.a$ are shown in square brackets. The blockchain indicator of activities is shown in the form of a chain symbol.
A solid chain represents on-chain tasks and data objects, while a dashed chain represents tracked activities and data objects where the digest is stored on-chain.
No chain symbol is used for off-chain data objects and activities.
For business rule tasks, the decision variable is shown in the upper right corner. General data flow between activities is shown graphically using the standard BPMN data flow edges between data objects and activities.

\subsection{Correctness of a Process Model}

\subsubsection{Correct Data Flow}
To provide correctness by design, a process model must guarantee that the inter-de\-pendencies of data flow and control flow are free of race conditions and that all read data objects are initialized. We ensure this property by the following criteria.
Let $I$ be an instantiation of decision variables of each xor-gateway and loop. $I$ contains a mapping between decision variables and Boolean values, whereas decisions in loops may contain an entry for each loop iteration. An instance type $P^I$ of a process $P$ is a sub-graph of $P$ where only the path of a gateway is followed, where the corresponding decision variable $I$ equals the condition of the branch, and loops are unrolled.
The origin of a data object $d$ for node $n$ with $d \in n.r$  in an instance type $P^I$, where $n \in P^I.N$ is the nearest predecessor node $w$ of $n$ in $P^I$ where $d \in w.w$.
The data flow of a process $P$ is correct, iff, for every possible instance type $P^I$, and for every reader of some data object, the origin is always unique.
This definition excludes typical data-flow flaws \cite{flawsflow} in processes like uninitialized variables and read-write dependencies or write-write dependencies of parallel executed tasks.

\subsubsection{Correct Blockchain-based Data-Flow}\label{correctbcdata}
The architecture of typical blockchain systems imposes strict restrictions on data flow between on-chain and off-chain nodes.
In addition, we aim to maximize the enforcement of the global model by the blockchain. 
We, therefore, require global processes to adhere to the following restrictions.

\paragraph{Data access of on-chain activities.}
Due to the typical limitations of blockchain systems, smart contract code can only read transaction input data and data already stored on-chain. 
On a blockchain, all transactions are recorded. This also includes every input parameter. 
Consequently, an off-chain data object gets permanently stored on-chain once used as an input parameter for an on-chain task.
Similarly, smart contracts can only write to on-chain data objects, resulting in the fact that all output data is implicitly stored on-chain.

Since we assume that designers carefully decide which data objects should be stored on-chain, we impose the following restriction:

\begin{Definition}[Correct data access of on-chain activities]\label{def:correct_bc_da}
The data access of on-chain activities of a global process model is correct, iff for every node $n \in P.N$ with $n.exec=`onChain`$, for every data object $d \in n.r \cup  n.w$;  $d.s=`onChain`$ holds.
\end{Definition}

This restriction does not limit the solution space of possible process implementations, as the designer can always introduce additional on-chain data objects and blockchain oracles to overcome this limitation. 

\paragraph{Data access of tracked activities}
A tracked activity is executed on an off-chain engine. However, the progress (start and end event) is tracked on the blockchain.
Such tasks can read and write to all data object storage types. The blockchain is in full control of write requests of blockchain data as the global execution state is stored on-chain. Also, for read requests, the blockchain can be efficiently used to guarantee that tracked data objects are only read by participants who are currently executing tasks having the tracked data object as input at runtime. 

\paragraph{Data access of off-chain activities}
An off-chain activity $a$ is executed in the process engine of its actor $a.a$. There are no global correctness guarantees since the blockchain cannot enforce the correct execution.
Our architecture aims to maximize blockchain-based enforcement and to guarantee that the control flow and data flow are exactly executed as modeled.
Therefore, off-chain tasks may not alter on-chain or tracked data since the blockchain cannot guarantee at runtime that a writing node is actually entitled to write based on the state of the global process. 
Consequently, off-chain tasks must only have off-chain data objects in their write-set.
This is different for read operations. Blockchain data is generally available for all participants. Thus, we assume that read operations on on-chain data are always possible for all participants. However, for off-chain activities, the blockchain has no information on the state of the process. Therefore, the blockchain cannot guarantee that an actor certainly needs tracked data. Consequently, off-chain tasks must not access tracked data.

\begin{Definition}[Correct data access of off-chain activities]\label{def:correct_offc_da}
The data access of off-chain activities of a global process model is correct, iff for every node $n \in P.N$ with $n.exec=`offChain`$, for every data object $d \in n.w$;  $d.s=`offChain`$ holds and for every data object $dr \in n.r$ $dr.s \in \{`onChain`,`offChain`\}$ holds.
\end{Definition}

\begin{figure}[H]
\center
\includegraphics[width=\textwidth]{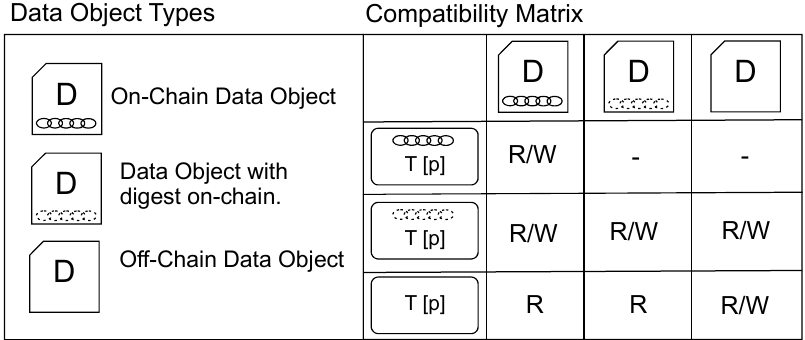}
\caption{Data Object Types and Compatibility Matrix}
\label{fig:dotypes}
\end{figure}

All requirements on the data flow of process models are summarized in Fig. \ref{fig:dotypes}.

\subsubsection{Correct Blockchain-based Control-Flow}
We follow a blockchain-first strategy. Therefore, if some state change of a process instance is recorded on the blockchain and some local participant process depends on this state update, the update is received from the blockchain and not by using direct message exchanges between the participants. The aim here is to maximize the degree of enforcement and to use the blockchain as a single source of truth whenever possible. 
However, at the same time, we assume that designers carefully decide what portions of the process should be executed on- or off-chain. Therefore, the blockchain should only store state changes of the process that are necessary according to the process model.

For xor-gateways, we follow the strategy that if the on-chain model requires a xor-gateway, the xor gateway must be executed on the blockchain. Thus, all dependent off-chain processes will read the state of the gateways from the blockchain.
However, if the model on the blockchain does not contain a specific gateway, off-chain communication is used to synchronize the runtime behavior of the gateway.
Parallel gateways and sub-processes/loops are treated based on the same principles. These assumptions result in the following constraints: 

\begin{Definition}[Correct blockchain-based control-flow]\label{def:Correct_blockchain-based_control_flow}
The blockchain-based control-flow of a correct global process model is correct if all following restrictions are met.
\begin{itemize}
\item For every xor-split node $x$ with the corresponding xor-join node $x'$: if there is a node $n$ with $n.exec=`onChain`$, $n.exec=`tracked`$, or $n.a=`chain`$ in one branch of $x$ and there exists another node $m$ with $m.exec=`onChain`$, $m.exec=`tracked`$ or $m.a=`chain`$ in the other branch of $x$ or anywhere outside of the block defined by $x$ and $x'$, then the actor of the $x$ and $x'$ must be the blockchain (actor $`chain`$). This requirement follows directly from the Augmentation rule for XOR in Def. \ref{def:xor_rule}.
\item  For every par-split node $p$ and the corresponding join node $p'$, the actor of $p$ and $p`$ must be $`chain`$, if there is an on-chain, tracked task, or node with actor $`chain`$ in both branches of $p$.
\item For every sub process $l$, the actor $l.a$ must be $chain$, if there is any on-chain or tracked activity or node with actor $`chain`$ in the body of $l$.
\end{itemize}
\end{Definition}
A model that does not comply with the requirements can always be transformed into a model complying with the requirements by changing the actor to $`chain`$, whenever required.

\section{Generating a Platform independent Augmentation} \label{sect:RulesPiA}
The global process model is first extended with the necessary steps for communication and synchronization between the different participants and the blockchain. We refer to this extended model as the Platform Independent Augmentation (PiA). A PiA of a process model $P$ is a process model $P'$ where the set of activity types is extended with the type $communicationTask$.
A node $c$ with type $communicationTask$ extends a node with the property $c.res \in P.A$. 
A communication task $c$ sends the set of data objects or decision variables $c.r$ from $c.a$ to actor $c.res$. A communication task with an empty read set only hands over the control flow to the recipient $c.res$. The graphical notation of communication tasks is borrowed from BPMN choreography tasks and shown in the lower left part of Figure \ref{GlobalprocessModel}.
Communication tasks are added to the process based on the following rules.

\subsection{GP to PiA Transformation Rules }\label{sec:abstract_rules}
\begin{figure}[H]
\center
\includegraphics[width=\textwidth]{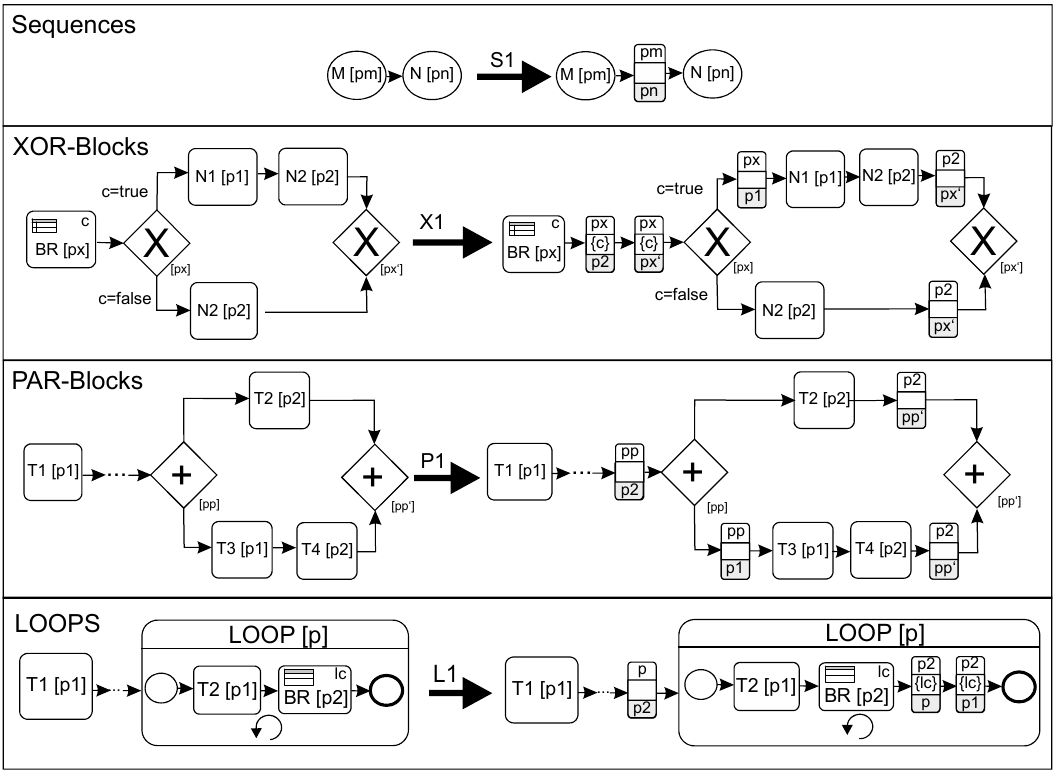}
\caption{Rules for the creation of the platform-independent augmentation}
\label{abstract_rules}
\end{figure}

We refer to the rules for deriving the PiA from the GP as abstract augmentation rules. They are based on previous work for the top-down development of inter-organizational processes in \cite{koepkeiiwas14}. The rules in \cite{koepkeiiwas14} have the benefit that their completeness and correctness are proven.
All rules are shown in Fig. \ref{abstract_rules} and defined below. Slight changes of the original rules in  \cite{koepkeiiwas14}  were needed since we use a graph-based process representation, while the original rules were defined on a block-based representation. 

\begin{Definition}[Sequence Rule S1]\label{def:seq_rule} 
If a different actor executes a successor node than its predecessor node, the actor of the predecessor hands over the control flow to the actor of the successor node via a communication step.\\
\\
\underline{Precondition:} There is an edge $(m,n,l) \in P.E$ where $m.a \neq n.a \wedge m.type \in \{`Activity`,$ $`Subprocess`$, $`XorJoin`$, $`ParJoin`\}$ $\wedge$
$n.type \in \{`Activity`,$ $`SubProcess`,$ $`XorSplit`,$ $`ParSplit`,$ $`SubProcess`,$ $`communicationTask`\}$ 
\\
\underline{Effects}: A new communication task $nc$ with $nc.a=m.a$, $nc.res=n.a$ is created. The edge $(m,n,l)$ is removed from $P.E$ and the set $\{(m,nc,l),(nc,n,l)\}$ is added to $P.E$.
The inserted abstract communication step hands over the control flow from the participant of node $m$ node to the participant of $n$.
\end{Definition}

\begin{Definition}[XOR-Rule X1]\label{def:xor_rule} 
Xor-split nodes have a business rule task that calculates their condition as their direct predecessor. The augmentation rule distributes the decision variable object to all dependent participants and adds the required control-flow handover via abstract communication steps.\\
\\
\underline{Preconditions:} There is a xor block defined by a $XorSplit$ node $xs$ and the corresponding $XorJoin$ node $xj$ in $P$. The direct predecessor of the xor-split node is a business rule task $b$ that writes the decision variable required by $xs$ \\
\\
\underline{Effects:} 
\begin{itemize}
\item A set of communication steps is inserted between $b$ and $xs$ to send the decision variable $c$ to all dependent participants. A participant $p$ is dependent of $xj$ if:
$p$ executes a node in both branches of the xor-block, and if $p$ executes a task in only one branch of the xor-block and anywhere outside of the xor-block.

\item If the executor of the $xs$ is not the executor of the following node $n$ in any branch, a communication task is inserted between $xs$ and $n$ handing over the control flow from $xs.a$ to $n.a$.
\item If the executor of the $xj$ is not the executor of the previous node $n$ in any branch, a communication task is inserted between $n$ and $xj$, handing over the control flow from $n.a$ to $xj.a$.
\end{itemize}
\end{Definition}

\begin{Definition}[PAR-Rule P1]\label{def:par_rule} 
For par blocks, the control flow needs to be passed between the split node and the first node in each branch and between the last node in each branch and the join node.\\
\\
\underline{Preconditions:}  There is a par block started by a $ParSplit$ node $ps$ and ended by the corresponding $ParJoin$ $pj$ node in $P$.\\
\\
\underline{Effects:} 
\begin{enumerate}
\item If the executor of the $ps$ is not the executor of the following node $n$ in any branch, a communication task is inserted between $ps$ and $n$ handing over the control flow from $ps.a$ to $n.a$.
\item If the executor of the $pj$ is not the executor of the previous node $n$ in any branch, a communication task is inserted between $n$ and $pj$, handing over the control flow from $n.a$ to $pj.a$.
\item For every actor $a$ who owns a node in both branches of the par-block and who owns no predecessor node of $ps$, a communication task from $ps.a$ to $a$ is added.
\end{enumerate}
\end{Definition}

Rule \ref{def:par_rule} is a slight variation of the original rule in \cite{koepkeiiwas14}. 
Effect 3 is not part of the original Rule. It was added to ensure that local processes where the first element is a parallel split gateway can be correctly started on BPMN engines.

\begin{Definition}[LOOP Rule L1]\label{def:loop_rule} 
For loops, two aspects need to be considered: First, the loop must be started at each dependent participant. This is implicitly the case for all participants who were actors of any previous node of the loop. For all others, a new process instance must be spawned, where the first element is the loop itself. This is realized by adding a communication step for control-flow handover from the loop actor to every participant who is part of the loop but not of any previous step. These communication steps are added directly before the loop task. The second aspect is the distribution of the decision variable of the Business Rule Task. This is achieved by adding communication steps that transmit the decision variable first to the loop actor (if different) and then to all actors within the loop.
\\
\underline{Preconditions:} There is a loop SubProcess $l$ in $P.N$ with the $BusinessRuleTask$ $b$ for computing the loop condition. \\
\underline{Effects:}  A set of communication tasks sending the decision outcome to all dependent participants is added immediately after $b$.
A participant depends on a loop $l$ if she is the actor of $l$ or an actor of any node in the loop body of $l$.
This allows each participant to re-iterate the loop sub-process locally.
As for $XORSplits$, additional communication steps are inserted directly before $l$ to allow all participants who are not executing any task before the loop to start the loop block locally. The actor of the loop is in charge of notifying all such participants about the loop execution.
\end{Definition}

We follow the principle of specific loop blocks of \cite{koepkeiiwas14} for loops. However, we show the rules for do-while loops rather than while loops here.

\section{Generating a Platform-specific Augmentation}\label{sec:genPisa}
The platform-independent augmentation does not consider what communication means are actually used and how the actors trigger the on-chain process model.
We derive a platform-specific augmentation by applying a set of transformation rules on the Platform-independent Augmentation. 

We follow a blockchain-first approach; therefore, if the on-chain process model needs to be informed about a state change, all dependent actors are also informed about the update via the blockchain.
We will now first introduce additional task types to realize the communication with the blockchain and off-chain components in Subsection \ref{sect:task_types}. We then introduce augmentation rules in Subsection \ref{sec:platform_specific_augnemtation_rules} to derive a platform-specific augmentation from an Platform independent Augmentation.

\subsection{Blockchain Specific Task Types}\label{sect:task_types}
We realize the communication between local processes and the on-chain processes by specific task types, which are either executed on-chain or off-chain. 
The graphical representation of the implementation-specific tasks is shown in the right part of Figure \ref{GlobalprocessModel}.

\begin{Definition}[Sign Execution Task]\label{def:sign_execution_task} 
A sign execution task is an off-chain task to be executed on some local process engine. A node $n$ with $n.type=`sign`$ has the additional property $n.sign$, containing a reference to an on-chain task that is triggered by the execution of $n$.
\end{Definition}

The name sign is used since this task creates a (signed) transaction and sends it to the blockchain to trigger the on-chain task. 

\begin{Definition}[Start Task]\label{def:start_task} 
A start task is used to track the start event of a tracked task on-chain. Therefore, a start task is itself an on-chain task. 
A start task $n$ has the $n.type=`StartT`$ and a reference to a tracked task $n.start$.
\end{Definition}

\begin{Definition}[End Task]\label{def:end_task} 
An end task is used to track the end event of a tracked task on-chain. Therefore, an end task is itself an on-chain task. 
An end task $n$ has the $n.type=`endT`$ and a reference to a tracked task $n.end$.
\end{Definition}

\begin{Definition}[Oracle Task]\label{def:oracle_task} 
An oracle task $o$ is an on-chain task with $o.type=`oracle`$ it stores all on-chain input data objects in $o.r$ on the blockchain.
\end{Definition}

\begin{Definition}[Reverse Oracle Task]\label{def:reverse_oracle_task} 
A reverse oracle task $ro$ is an off-chain task with $ro.type=`revOracle`$. It is used to read data objects from the chain and to get notified about the control flow of the on-chain process. All data objects of the $ro.r$ are read from the chain and provided as local data objects to the calling participants. A reverse oracle task has the additional property $ro.waitFor$ that refers to an on-chain node. It only returns after the completion event of $ro.waitFor$ is recorded on-chain relative to the current loop iteration.  
\end{Definition}

\subsection{Local Data Objects}\label{sect:local_data objects}
In the platform-independent augmentation, each data object can be accessed by all tasks with the data objects in their read or write sets, and no data objects for transporting data elements between on- and off-chain components are introduced. The platform-specific augmentation now addresses the case that a local engine can only access local data objects. 
Platform-specific Augmentations therefore include local data objects. Such data objects have an additional type attribute with the filler $`local`$.
Each such data object has the properties $a$ $\in$ $P.A$, $d \in \{ do \in P.D : do.type \neq `local`\}$.
$a$ defines the participant who is entitled to use the data object locally. The property $d$ refers to the global data object that is replicated in the local data object. For local data objects, $d.d$, and $d.a$ together form a key in the sense that two data objects with the same values for $d.d$, and $t.a$ refer to the same object.

For digest on-chain data, participants need to keep track of both a local copy of the data using a local data object and a data object for storing the hash value from the chain locally. This is realized by data objects with the type $`hash`$. Such data objects have the same properties and key properties as local data objects.

\subsection{Control Flow Communication Rules} \label{sec:platform_specific_augnemtation_rules}

\subsubsection{Rules for control flow handover}

\begin{figure}[H]
\center
\includegraphics[width=\textwidth]{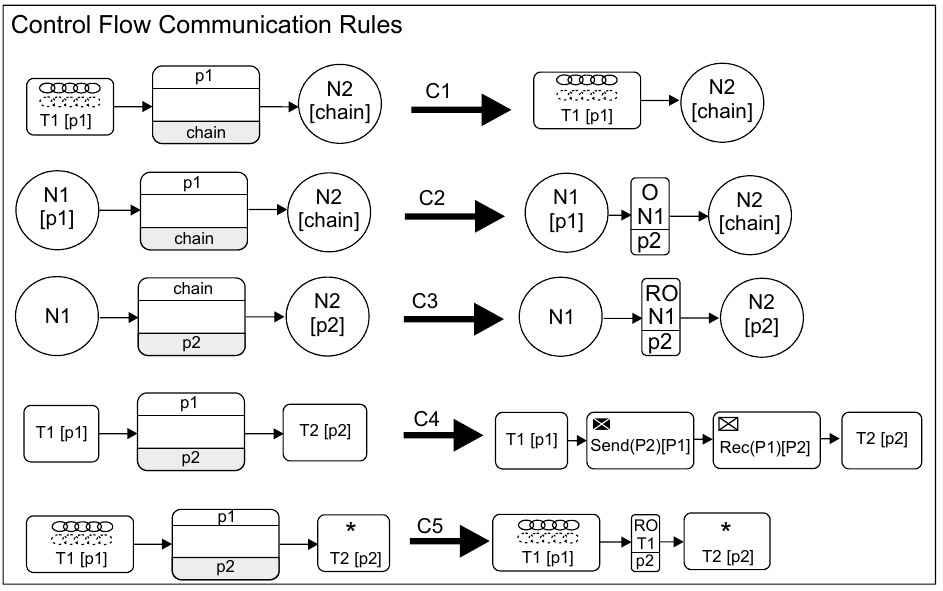}
\caption{PiA to PsA transformation of the control-flow}
\label{fig:nodereplacement}
\end{figure}
\begin{Definition}[Communication Task onChain or tracked to chain C1]\label{def:Rule_C1} 
If an abstract communication step sends the control flow from an on-chain or tracked task to a node with an actor $`chain`$, then the communication step can be removed as the state is fully known on the blockchain.\\
\\
\underline{Preconditions:}
$\exists c \in P.N : c.type=`CommunicationTask` \wedge c.a \neq `chain` \wedge c.r = \{\}$ \\
$\wedge \exists (x,c) \in P.E : (x.exec=`chain` \vee x.exec=`tracked`) \wedge c.res=`chain`$\\
$\wedge \exists (c,y) \in P.E : y.a=chain$\\
\\
\underline{Effects:} $(x,c,\_)$ and  $(c,y,\_)$ are removed from $P.E$, $c$ is removed from $P.N$, new edge $(x,y,_)$ is added to $P.E$
\end{Definition}

\begin{Definition}[Communication Off-Chain to On-Chain C2]\label{def:Rule_C2} 
There is a communication step with $chain$ as a receiver, the previous step is not an on-chain or tracked task. In this case, the communication task is replaced by an oracle task. Therefore, the respective off-chain engine notifies the blockchain about the state update.\\
\\
\underline{Precondition:}
$\exists c \in P.N : c.type=`CommunicationTask` \wedge c.a \neq `chain` \wedge c.res = `chain` \wedge c.r = \{\}$ 
$\wedge \exists (x,c,\_) \in P.E : x.a \neq `chain` \wedge \exists (c,y,\_) \in P.E : y.a = `chain`$\\
\\
\underline{Effects:}
A new oracle task $o$ with $o.a = c.a$  is created. $c$ is replaced by $o$.
\end{Definition}

\begin{Definition}[Communication Task Chain to Off-Chain C3]\label{def:Rule_C3} 
There is a communication Task with $chain$ as actor, the previous task has actor chain or is on-Chain and the next task does not have chain as actor.  In this case, the communication is replaced by a reverse oracle. Therefore, the off-chain engine waits for the previous on-chain node to be executed.\\
\\
\underline{Precondition:}
$\exists c \in P.N : c.type=`CommunicationTask` \wedge c.a = `chain` \wedge c.r = \{\}$ \\
$\wedge \exists (x,c.\_) \in P.E : x.a=chain \wedge c.res \neq chain$
$\wedge \exists (c,y.\_) \in P.E : y.a \neq chain$\\
\\
\underline{Effects:}
A new reverse oracle task $ro$ with $ro.a = c.res$ and $ro.waitFor=x$ is created. $c$ is replaced by $ro$.
\end{Definition}

\begin{Definition}[Communication Off-Chain to Off-Chain C4]\label{def:Rule_C4} 
There is a communication step $c$ where $c.a \neq `chain`$ and $c.res \neq `chain`$ and the previous task is off-chain and the next task of off-chain. Then the communication is realized via a send task of $c.a$ and a receive task of c.res.\\
\\
\underline{Precondition:}
$\exists c \in P.N : c.type=`CommunicationTask` \wedge c.a \neq `chain` \wedge c.res \neq `chain` \wedge c.r = \{\}$ 
$\wedge \exists (x,c.l1) \in P.E : x.exec = `offChain` \wedge \exists (c,y.l2) \in P.E : y.exec = `offChain`$\\
\\
\underline{Effects:}
a new task $s$ with $s.type=`send`$ and $s.a =c.a$ and a new task $r$ with $r.type=`receive`$ and $r.a=c.res$ are created.   
In $P.E$ $(x,c,l1)$ is replaced by $(x,s,l1)$, $(s,r,l1)$ is added and $(c,y,l2)$ is replaced by $(r,y,l2)$.
\end{Definition}

\begin{Definition}[Communication Off-Chain to Off-Chain after on-chain or tracked task  C5]\label{def:Rule_C5} 
There is a communication task $c$ where $c.a \neq `chain`$ and $c.res \neq `chain`$ and the previous task is an on-chain or tracked task. Then the communication is realized via a reverse oracle notifying the next task about the completion of the previous tracked or on-chain task.\\
\\
\underline{Precondition:}
$\exists c \in P.N : c.type=`CommunicationTask` \wedge c.a \neq `chain` \wedge c.res \neq `chain` \wedge c.r = \{\}$ 
$\wedge \exists (x,c,l1) \in P.E : x.exec \neq `offChain` \wedge \exists (c,y,l2) \in P.E$\\
\\
\underline{Effects:}
a new task $ro$ with $ro.type=`revOracle`$ and $ro.a =c.res$ $ro.waitFor=x$ is created.   
In $P.E$ $(x,c,l1)$ is replaced by $(x,ro,l1)$, $(c,y,l2)$ is replaced by $(ro,y,l2)$.
\end{Definition}

\subsubsection{Rules for distributing decision outcome}

\begin{figure}[H]
\center
\includegraphics[width=\textwidth]{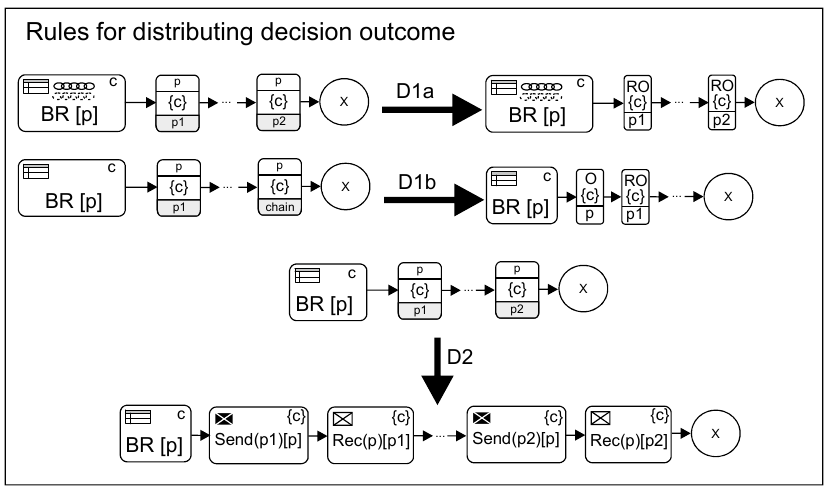}
\caption{Rules for distributing decision outcome for PsA}
\label{fig:rules_control_flow_decisions}
\end{figure}

While the previous rules excluded communication tasks transmitting decision variable, we now provide rules specifically for distributing decision variable to support xor and loops. 

\begin{Definition}[Transmitting decision outcome via the Blockchain D1a]\label{def:Rule_D1a} 
There is a sequence of tasks: $br, c_1, c_2, ..., c_n, x$ in $P$ where each $c$ transmits a decision variable $d$ written by $br$, and $x$ is not a communication step sending $d$.
If $br.exec=`onChain`$ or $br.exec=`tracked`$ then the decision is written to the blockchain.
Therefore, any communication task in $c_1-c_n$ with $c.res=`chain`$ is removed, and all others are replaced by reverse oracles reading the respective variable.
\end{Definition}

The previous rule covers the case that the decision variable is stored on-chain because the business rule task is on-chain or tracked. Otherwise, if the actor on-chain is dependent but the business rule task is not executed on-chain or tracked, the data object must be stored on-chain via an oracle:

\begin{Definition}[Transmitting decision outcome via the Blockchain D1b]\label{def:Rule_D1b} 
There is a sequence of tasks: $br, c_1, c_2, ..., c_n, x$ in $P$ where each $c$ transmits a decision variable $d$ written by the $BusinessRuleTask$ $br$, with $br.exec=`offChain`$ and $x$ is not a communication step sending $d$.
If there is at least one $c$ in $c_1-c_n$ with $c.res=`chain`$, then this communication step is removed, an oracle task storing $d$ on-chain is placed immediately after $br$, and all others are replaced by reverse oracles.
\end{Definition}

\begin{Definition}[Transmitting decision outcome via messages D2]\label{def:Rule_D2} 
There is a sequence of tasks: $br, c_1, c_2, ..., c_n, x$ in $P$ where each $c$ transmits a decision variable $d$ written by the $BusinessRuleTask$ $br$, and $x$ is not a communication step sending $d$.
If there exists no communication step $c$ in $c_1-c_n$ with $chain$ as actor or receiver, $d.s \neq `chain`$, then the decision is distributed via messages to all recipients.
Each communication task is replaced by a send-task of $c.a$ and a receive step with actor $c.res$.
\end{Definition}

\subsection{Examples of Rule Application}

\begin{figure}[H]
\center
\includegraphics[width=\textwidth]{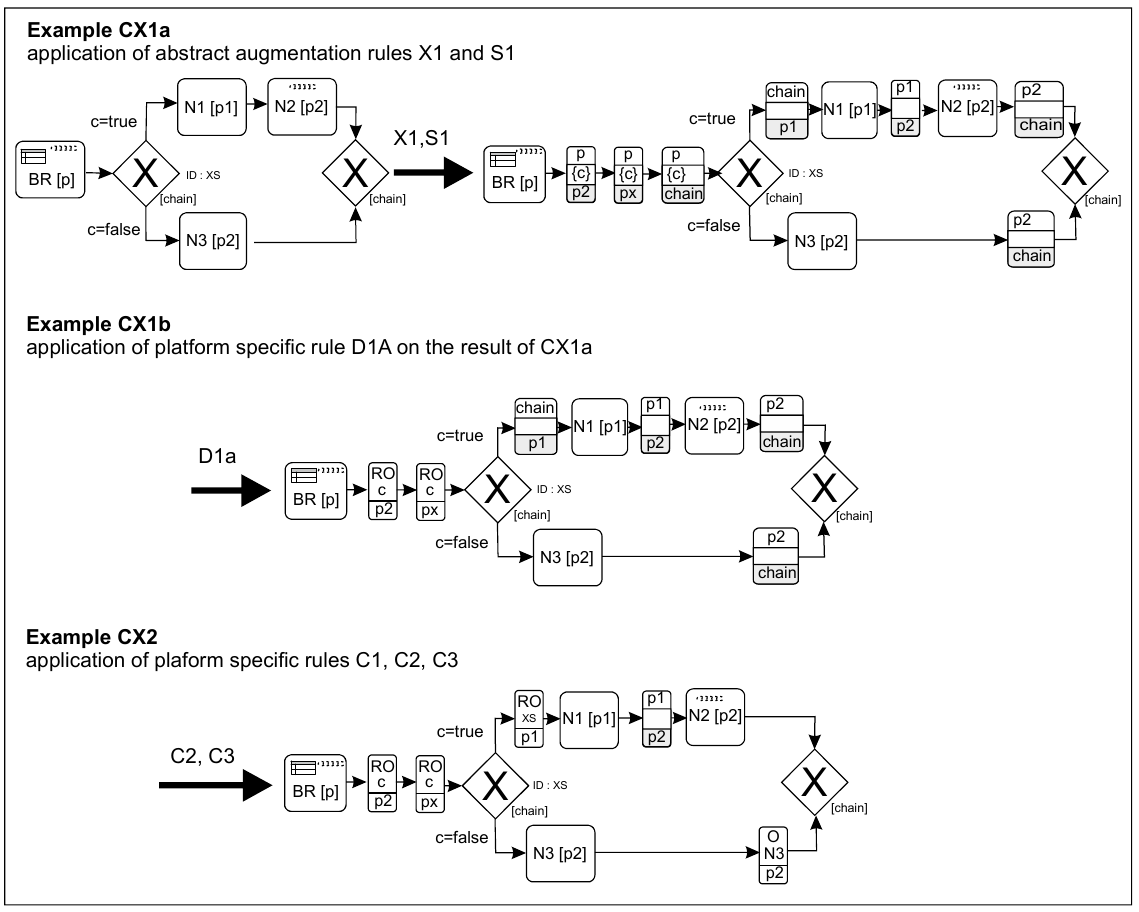}
\caption{Examples of the combined rule application for XOR-Blocks}
\label{fig:example_composite_xor}
\end{figure}

We will now provide examples of the interplay of abstract and platform-specific rules. The cases are illustrated in Figure \ref{fig:example_composite_xor}.
Example CX1a shows the input and result of applying the abstract augmentation rules X1 and S1. These rules are independent of any blockchain properties.

In Example CX1b, the blockchain-specific rule for the distribution of the decision outcome is applied (D1a). The Business Rule task is a tracked task. Therefore, the decision outcome is implicitly stored in the blockchain. Accordingly, D1A rewrites the communication steps transferring c to a reverse oracle and any communication step where $chain$ is the recipient is removed. For clarity and a compact representation, we do not apply the rule C4 in this example.

Subsequently, the rules $C1$, $C2$ and $C3$ are applied on the result of $CX1b$ in Example $CX2$. 
$C1$ removes the non-required communication strep between the tracked task $N2$ and the xor-join node.
$C2$ Rewrites the existing communication step between $N3$ and the xor-join gateway to an oracle. In this case, $p2$ has to notify the on-chain process about the completion of the off-chain task $N3$. $C3$ rewrites the communication step between the xor-split and $N1$ to a reverse oracle.

\begin{figure}[H]
\center
\includegraphics[width=\textwidth]{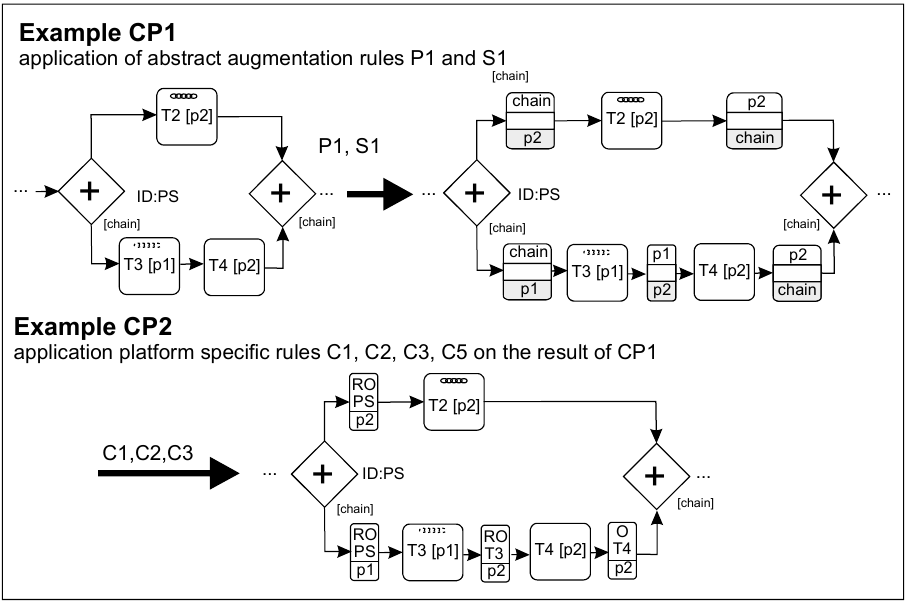}
\caption{Examples of the combined rule application for PAR-Blocks}
\label{fig:example_composite_par}
\end{figure}

Figure \ref{fig:example_composite_par} shows a combined example of the rule application for par blocks. 
Example $CP1$ shows the abstract augmentation derived by rules $P1$ and $S1$ for a fragment of a process with a parallel block. 
Example $CP2$ shows the transformation of the result of $CP1$ to a platform-specific augmentation using $C1$, $C2$, and $C3$.
In particular, rule $C1$ removes the Communication step between $T2$ and the parallel join. This is possible since $T2$ is an on-chain task. Therefore, the chain is already aware of the completion of $T2$. Rule $C2$ translates the communication step between $T4$ and the parallel join node to an oracle, since $T4$ is an off-chain task and therefore, $P2$ has to notify the on-chain process about the completion of $T4$. Rule $C3$ is used to replace the communication step between the parallel split and $T2$ and $T3$ by Revers Oracles. Therefore, $p2$ and $p1$ are notified via the blockchain about the execution of the parallel split. Finally, $C5$ is applied to hand over the control-flow between $T3$ and $T4$ via the blockchain. This is beneficial since $T3$ is a tracked task.

\begin{figure}[H]
\center
\includegraphics[width=\textwidth]{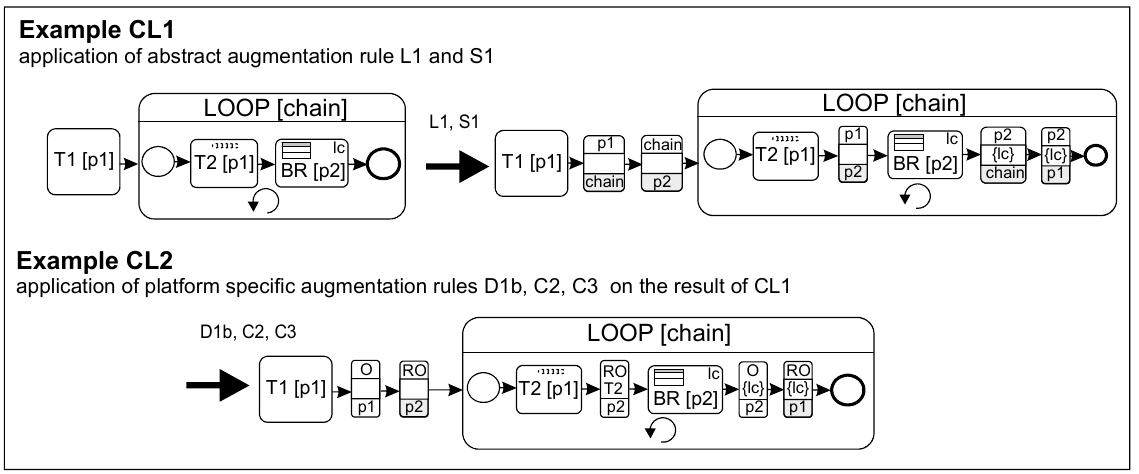}
\caption{Examples of the combined rule application for Loops}
\label{fig:example_composite_loops}
\end{figure}

Figure \ref{fig:example_composite_loops} shows the processing of loop blocks.
Example $CL1$ shows the application of the abstract rules $L1$ and $S1$. The first applied $L1$ rule adds a communication step from the actor of the loop sub-process $chain$ to $p2$ in order to allow $p2$ to start the loop sub-process locally. In addition, $L1$ adds communication steps after the business-rule task to all dependent actors ($chain$,$p1$) that transmit the decision outcome in variable $lc$.
Next, $S1$ is executed. It leads to the inclusion of a communication step for control-flow hand-over after task $T1$ between $p1$ and $chain$. This allows the actor of the loop sub-process $chain$ to start the execution of the loop. In addition, $S1$ leads to the inclusion of a communication step after $T2$ for handing over the control-flow to $P2$ for starting $BR$.

In the following (Examples CL2) the rules $D1b$, $C2$ and $C3$ are applied.
$D1b$ rewrites the existing communication steps for distributing the decision variable $lc$ to an oracle executed by $p2$ and a reverse oracle executed by $p1$.
Next, $C2$ replaces the communication-step after $T1$, transmitting to $chain$ by an oracle. 
$C3$ replaces the communication step immediately before the loop sub-process that sends from $chain$ to $p$2 by a reverse oracle.

\subsection{Rules for Starting Tasks} \label{task-rules}

\begin{figure}[H]
\center
\includegraphics[width=\textwidth]{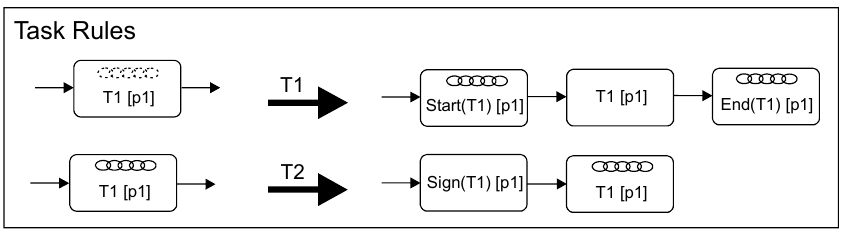}
\caption{Rules for starting tasks on the blockchain}
\label{fig:task_rules}
\end{figure}

By design, blockchain systems can only execute code if a corresponding transaction triggering the code execution is sent from some off-chain component to the blockchain.
In our architecture, we include the required calls creating the transactions in the off-chain processes of the participants. This is realized by the following rules:

\begin{Definition}[Tracked Task T1]\label{def:Rule_T1} 
A tracked task is a task that is executed on the local process engine of some participant. However, the start and completion events are recorded on-chain.
For every tracked task $t$ ($t.exec=tracked$), we generate a start task $st$ with $st.type=StartT$ and $st.start=t$ and
an end task $e$ with $e.type=EndT$ and $e.end=t$. 
We replace every edge $(\_,t,l1)$ in $P.E$ by $(\_,st,l1)$ and add the edge $(st,t,l1)$, and replace every edge $(t,\_,l2)$ by  $(e,\_,l2)$ and add the edge $(t,e,l1)$. Therefore, every tracked task $t$ is replaced by a sequence $<start$ $t$,$ t,$ $end$ $t>$.
Start and end tasks are on-chain tasks.
\end{Definition}

\begin{Definition}[On-Chain Task T2]\label{def:Rule_T2} 
For every onChain Task $t$ ($t.exec=onChain$) and $t.a \neq chain$ we generate a sign execution task $se$ with $se.type=sign$ and $se.start=t$.
We replace every edge $(\_,t,l)$ in $P.E$ by $(\_,se,l)$ and add the edge $(se,t,l)$.
\end{Definition}

\section{Extending PsA with Data Access}\label{sect:dataAccess}
After the previous augmentation rules are executed, the control flow of the model is already correctly realized. However, the data flow between activities reading and writing general data objects does not exist yet. We now discuss the possible cases of data access to motivate the chosen realization within PsA's.

\subsection{On-chain tasks accessing on-chain data objects}
Each on-chain task is meant to be realized by a custom smart contract function / transaction.
This function needs to be realized in a later phase. However, since on-chain code can generally access on-chain data (typically except read access to log data),
no further processing of the process model is required for this case. The tasks to be implemented in the form of smart contracts and their data access behavior can be derived from the BPMN model. No further processing of the model itself is required.

\subsection{Tracked task accessing on-chain data objects}
Tracked tasks are executed by a local process engine, and only their start and completion events are recorded on-chain using blockchain transactions.
According to the rules in Section \ref{task-rules}, for a tracked task, first, the off-chain engine executes a sign step for an on-chain start task.
This causes the execution of the on-chain start task. Then, the off-chain engine executes the task off-chain and reports its completion by signing for the end task, which results in the execution of the end task on-chain.

Since we aim for minimal changes in process engines, we use the sign task to implement read and write operations of on-chain data objects.
For reads, the sign task of the start task provides all on-chain input data objects to the off-chain engine, while the sign task of the end step writes the respective local data objects to the chain. In both cases, this is realized by the interplay of the sign task and the corresponding start and end tasks that implement the on-chain part.

\begin{figure}[H]
\center
\includegraphics[width=\textwidth]{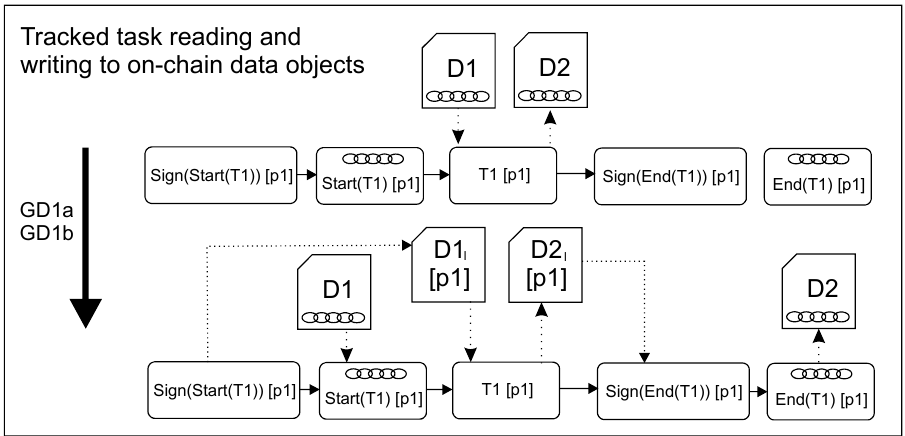}
\caption{GD1: Tracked Tasks accessing on-chain data objects}
\label{fig:do_rules_cropped.pdf}
\end{figure}

The rules for realizing on-chain data access of tracked tasks are illustrated in Fig. \ref{fig:do_rules_cropped.pdf} and defined below.\\

\begin{Definition}[Reading of on-chain data (GD1a)]\label{def:GD1a} 
For every tracked task $t$ and every on-chain data object $d$ in the read set of $t$, a local data object $d_l$ with $d_l.type=`local`$, $d_l.a=t.a$, $d_l.d=d$ is created. $d_l$ is added to the write set of the sign step of the start task of $t$, while $d$ is added to the read set of the start task of $t$. Data Object $d$ is removed from the read set of $t$ and added to the read set of the start task of $t$.\\
\end{Definition}

\begin{Definition}[Writing of on-chain data (GD1b)]\label{def:GD1b} 
For every tracked task $t$ and every on-chain data object $d$ in the write set of $t$, a local data object $d_l$ with $d_l.type=`local`$, $d_l.a=t.a$, $d_l.d=d$ is created. 
$d_l$ is added to the write set of $t$ and the read set of the sign-task of the end task of $t$. Finally, $d$ is removed from the write set of $t$.
The original on-chain data object $d$ is added to the write set of the on-chain end task of $t$ and removed from the one of $t$.
\end{Definition}

\subsection{Tracked task accessing digest on-chain data objects}
Data objects with storage type digest are stored locally at each participant. Only the hash value of the data object is stored on-chain together with the participant who last updated the data. For reading such a data object, it is therefore not sufficient to read the hash value from the chain. Instead, the actual data value needs to be retrieved from the participant with the most recent version of the data object.  
To do this, we introduce an additional task of type Data helper (DH). Such a task receives a set of pairs of hash values and a participant as input and collects the required data objects from each participant.  

We assume that if a data object has the type digest on-chain, the data needs to be kept confidential. Therefore, only participants who certainly need a data value for executing tasks should be able to retrieve the clear text data. Therefore, a data value should only be sent to a participant if they certainly execute a task needing that data value. We achieve this by the following architecture.

\subsubsection{Data handling architecture}
Each participant provides a component for handling incoming read requests of data objects with digest on the chain. This component accepts data requests from the other participants. Tracked tasks that need a data object with digest on-chain must first execute the start task for the task to be executed. This allows them to retrieve the hash value and the originating participant from the chain, and it records the fact on chain that they are executing the task. 
They then send a request to the data handling service of the last writer of the data object. This request contains the participant's identity, the ID of the process instance, and the hash value of the data object. The data handling service only replies with the value of the data object if both of the following conditions apply:

\begin{itemize}
\item The hash value of the receiver's data object for the given process ID matches the hash provided in the request, and the hash matches the most recent on-chain hash value. This ensures that the receiver actually has the required data for process execution.
\item The execution of the start task reading that data object was recorded on-chain for that process instance, and the end task was not recorded yet. In the case of loops, this holds only for the current loop iteration. This ensures that only participants who execute tasks needing the data object can obtain the data value. 
\end{itemize}

\subsubsection{Augmentation rules for digest on-chain data}
\begin{figure}[H]
\center
\includegraphics[width=\textwidth]{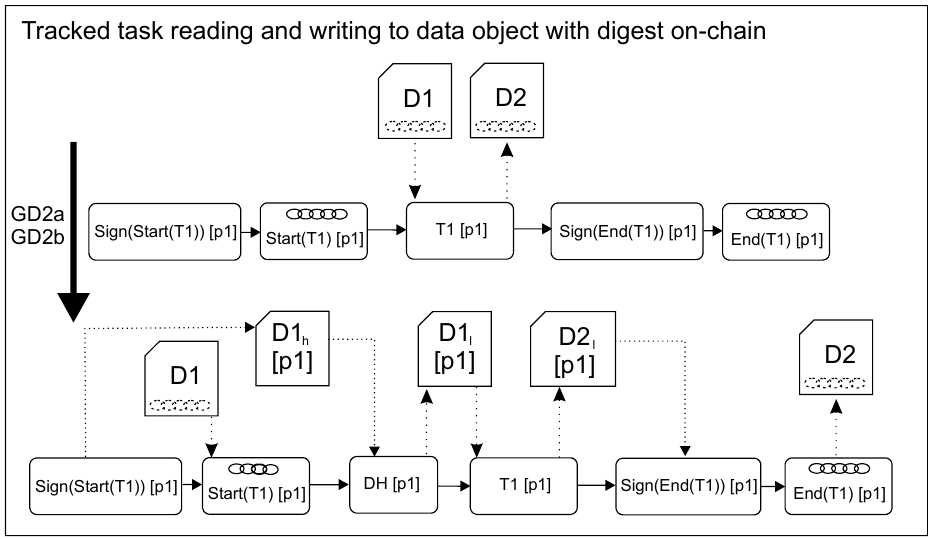}
\caption{GD2: Tracked Tasks accessing data with digest on-chain}
\label{fig:do_rules_croppedDigest.pdf}
\end{figure}

Based on the described architecture, we use the following augmentation rules for reading and writing to data objects with digest on-chain.
The rules for read and write access are illustrated in Figure \ref{fig:do_rules_croppedDigest.pdf} and defined below.

\begin{Definition}[Reading of data objects with digest on chain (DG2a)]\label{def:GD2a} 
For every tracked task $t$ and every data object with digest on chain $d$ in the read set of $t$, two local data objects $d_h$ and $d_l$ with 
$d_h.type=`hash`$, $d_h.a=t.a$, $d_h.d=d$, and  $d_l.type=`local`$, $d_l.a=t.a$, $d_l.d=d$ are created. Data object $d_h$ is used for temporal storage of the on-chain hash value, and the last writing participant of the data, $d_l$, is used to store the actual data value locally.

The data object $d_h$ is added to the write set of the sign task of the start task of $t$. The on-chain data object $d$ itself is added to the read-set of the start task of $t$. A new task $DH$ of type data handling task that is used to send requests to the data remote handling component is added if it does not yet exist immediately after the start task. The task is an off-chain task and is assigned to $t.a$. The data object $d_h$ is added to the read set of $DH$. $d_l$ is added to the write set of $DH$. Finally, the data object $d$ in the read set of $t$ is replaced by $d_l$.
\end{Definition}

\begin{Definition}[Writing of data objects with digest on chain (GD2b)]\label{def:GD2b} 
For every tracked task $t$ and every data object $d$ with digest on chain in the write set of $t$, a local data object $d_l$ with $d_l.type=`local`$, $d_l.a=t.a$, $d_l.d=d$ is created. $d_l$ is added to the write set of $t$ and the read set of the sign-task of the end task of $t$. Finally, $d$ is removed from the write set of $t$. The original on-chain data object $d$ is added to the write set of the on-chain end task of $t$ and removed from the one of $t$. Here, the assumption is that the sign task reads $d_l$, computes the hash value of it, and uses is as transaction input for the end transaction sent to the blockchain. The on-chain end task will actually update the value on the chain. 
\end{Definition}

\subsection{Off-chain tasks reading on-chain data objects}
\begin{figure}[H]
\center
\includegraphics[width=\textwidth]{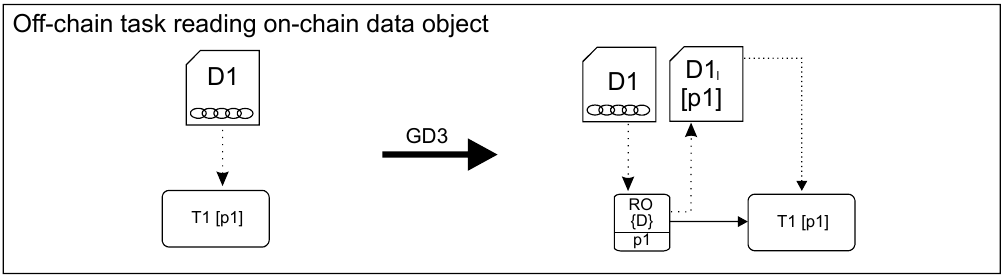}
\caption{GD3: off-chain task reading on-chain data}
\label{fig:GD3}
\end{figure}

As on-chain data objects are available to all process participants at any time, off-chain tasks are allowed to read on-chain data objects.
This is realized by adding a reverse oracle task $ro$ directly before the off-chain task. The task $ro$ has the on-chain data object in its read set and a local copy of it in its write set. The off-chain task itself reads the data from the local copy.
The approach is illustrated in Fig. \ref{fig:GD3}, the rule is defined below

\begin{Definition}[Off-Chain tasks reading on-chain data objects (GD3)]\label{def:GD3} 
For every off-chain task $t$ with some on-chain data objects in its read set, a new reverse oracle task $ro$ with $ro.a = t.a$ is created.
$ro.r = \{dx \in t.r : dx.s = `chain` \}$. Each edge $(x,t,l)$ in P.E is replaced by $(x,ro,l)$, and the edge $(ro,t,l)$ is added to $P.E$. 
\end{Definition}

\subsection{Off-Chain tasks reading and writing off-chain data objects}
When a node $m$ writes to a data object $d$ in an instance type and later a node $n$ with $m$ as the origin for $d$ reads the data object, the data object must be available for the actor of $n$. In the case of off-chain data objects, we realize the data flow via standard messages sent by send-and-receive tasks in the process model. Here we use the "Early Send" strategy \cite{Köpke2019}. Therefore, after any node $w$ that writes to an off-chain data object $d$ ($d \in w.w \wedge d.s=`offChain`$), the actor of $w$, $w.a$ sends the new data value to all participants, where $w$ is a possible origin. Since the definition of origins in  \cite{Köpke2019} did not cover loops, we extend the approach to loops here.

To calculate all peers who may have $w$ as their origin for a data object $d$ in the presence of loops, we use the following approach:
A temporary model $P'$ of $P$ is created, where all loop blocks are replaced by a sequence produced by including the loop body two times in the process model.
The second iteration is nested into a xor-block. Now, for all writing nodes $w$ in the first loop iteration, we compute all nodes where $w$ is a possible origin. After that, the corresponding communication steps to transmit the data objects to all potential readers are added to the original model.

In this paper, we focus on the blockchain-based implementations of processes. Depending on the actual requirements, other patterns for the implementation of the data flow can be used. We refer the interested reader to \cite{Köpke2019} for all the details. However, sophisticated and optimized patterns for off-chain data access are challenging, especially in the presence of loops, since in the off-chain case, no global state is available. 

\subsection{Rule Application Order} \label{sect:RuleApplicationOrder}
The rules are, by purpose, dependent on each other. The outcome of the GP to PiA rules is the input for the PiA to PsA rules.

\begin{enumerate}
    \item\textbf{GP to PiA Rules} \\
     L1 (Def. \ref{def:loop_rule}), X1 (Def. \ref{def:xor_rule}), P1 (Def. \ref{def:par_rule}), S1 (Def.\ref{def:seq_rule}).
    \item{\textbf{PiA to PsA Rules}\\
         Decision Distribution Rules:\\
         D1a (Def. \ref{def:Rule_D1a}), D1b (Def. \ref{def:Rule_D1b}), D2 (Def. \ref{def:Rule_D2})\\
         Control-Flow Rules:\\
        C1 (Def. \ref{def:Rule_C1}), C2 (Def. \ref{def:Rule_C2}), C3 (Def. \ref{def:Rule_C3}), C4 (Def. \ref{def:Rule_C4}), C5 (Def. \ref{def:Rule_C5})\\
        Task Rules:\\
        T1 (Def. \ref{def:Rule_T1}), T2 (Def. \ref{def:Rule_T2})\\
        Data-Access Rules:\\
        GD1a (Def. \ref{def:GD1a}), GD1b (Def. \ref{def:GD1b}), GD2a (Def. \ref{def:GD2a}), GD2b (Def. \ref{def:GD2b}), GD3 (Def. \ref{def:GD3})\\
        }
\end{enumerate}

\section{Generating local and blockchain Models from PsA's}\label{sect:projection}

The last step is to generate one local Process Model for each participant and one local model for the blockchain from the PsA.
This is achieved using the projection approach presented in  \cite{koepkeiiwas14}.
Basically, a task is only projected to the process model of the actor of that task. An on-chain task is projected only to the on-chain model.
Gateways are projected to all dependent participants. Gateways with actor chain are projected to the on-chain model. 
Control-flow edges of the PsA are projected to a specific target model if both nodes appear in the target model. Otherwise, new control-flow edges are introduced to connect nodes locally in the target models that are transitively connected in the PsA. We refer the interested reader to \cite{koepkeiiwas14} for all details. While the control-flow projection is directly adopted from \cite{koepkeiiwas14}, new projection rules are required for the implementation of the data access. 

\paragraph{Projection of Data Objects}
Data objects of type local and hash are only projected to the respective participant.
Data objects of type digest or on-chain are projected to every model containing at least one node where they are in the read or write set.

\section{Conclusion}
In this paper, we have presented a set of rules for generating local process models and an on-chain model automatically from a global process model. The interplay of the local models and the blockchain model ensures that the global process model is faithfully implemented. This approach facilitates the modeling and execution of blockchain-based collaborations in a top-down manner: 
First, participating organizations agree on a global process model that defines the activities that need to be performed by each participant. Next, designers can specify the tasks and data objects that require blockchain support and define the degree of support required for each. This allows the implementation of various security requirements of the global process. 
Our rules enable the automatic generation of the local and blockchain models. Local models can be executed using standard process engines like Camunda BPMN. The blockchain model, on the other hand, can be implemented via generated smart contracts or executed using a blockchain-based process execution engine.

\bibliographystyle{spmpsci}      
\bibliography{literatur}   

\end{document}